# ON CLIENT'S INTERACTIVE BEHAVIOUR TO DESIGN PEER SELECTION POLICIES FOR BITTORRENT-LIKE PROTOCOLS

Marcus V. M. Rocha[1] and Carlo Kleber da S. Rodrigues[2,3]

[1]Minas Gerais State Assembly, MG, Brazil
`marcus@melorocha.org`
[2]Electrical and Electronics Department, Polytechnic School of Army, Sangolquí, Ecuador
[3]University Center of Brasilia, Brasília, DF, Brazil
`carlokleber@gmail.com`

## ABSTRACT

*Peer-to-peer swarming protocols have been proven to be very efficient for content replication over Internet. This fact has certainly motivated proposals to adapt these protocols to meet the requirements of on-demand streaming system. The vast majority of these proposals focus on modifying the piece and peer selection policies, respectively, of the original protocols. Nonetheless, it is true that more attention has often been given to the piece selection policy rather than to the peer selection policy. Within this context, this article proposes a simple algorithm to be used as basis for peer selection policies of BitTorrent-like protocols, considering interactive scenarios. To this end, we analyze the client's interactive behaviour when accessing real multimedia systems. This analysis consists of looking into workloads of real content providers and assessing three important metrics, namely temporal dispersion, spatial dispersion and object position popularity. These metrics are then used as the main guidelines for writing the algorithm. To the best of our knowledge, this is the first time that the client's interactive behaviour is specially considered to derive an algorithm for peer selection policies. Finally, the conclusion of this article is drawn with key challenges and possible future work in this research field.*

## 1. INTRODUCTION

On-demand streaming systems ideally let their clients select multimedia objects, remotely stored in a server or group of servers, to be played at any instant of time with a satisfactory quality of service (QoS), mainly observed in terms of high playout continuity and low latency. The interactive access, by its turn, allows the clients to execute DVD-like actions during the object playback. These types of systems have become extraordinarily popular due to the recent and permanent advances in the field of network technologies, especially those resulting in more bandwidth capacity, reliability, confidentiality and scalability [1, 2].

For implementing on-demand streaming systems, the peer-to-peer (P2P) swarming architecture has already been proven to be a more effective solution than the classical client-server architecture, even if deploying content distribution networks (CDN) and IP multicast [1, 3]. The swarming architecture dictates that each peer is directly involved in the object delivery and contributes with its own resources to the streaming session (i.e. *perpendicular bandwidth*), thus provoking a load balance over the entire network and making the system scale well. Hence, increasing the peer population not only increases the workload, but also produces a concomitant increase in service capacity to process the request workload.

The client-server architecture, by its turn, deals with serious limitations due to the bandwidth bottleneck at the server or group of servers from which all clients request the objects. That is, an increase in the client population also increases the workload, inducing the scalability problem and drastically degrading system performance [3, 4, 5, 6].

P2P swarming protocols are essentially based on the definition of two main policies: piece selection and peer selection. The former refers to how the pieces of a multimedia object are going to be chosen by a peer (client, user, node, etc.) for download, while the latter dictates to which peers the pieces of the object should be uploaded to [3]. These two policies must stimulate the cooperation and reciprocation between peers to make the system scale well. With this in mind, there have been studies focusing on incentive politics to encourage peer contribution and to guarantee service fairness as well [3, 4, 5, 6]. Besides, the growing recognition of reputation systems [7] has generated significant research in this field too. Reputation-based schemes determine service priorities based on the histories of peer contributions to the system: reputable peers receive high priority and are rewarded more resources to compensate for their eventual contributions [8, 5].

The great efficiency of P2P swarming protocols for object (content) replication has surely attracted a lot of interests from academia and industry, and has notably motivated a number of proposals to adapt these protocols to meet the requirements of on-demand streaming systems. In general, these requirements refer to satisfying time constraints, namely: the client wants to play the object as soon as he arrives to the system and does not tolerate any discontinuities during the object playout [3, 9, 10].

The vast majority of these proposals of adaptation consists of modifying the piece selection and peer selection policies, respectively, of the original P2P swarming protocols. Nevertheless, the research community has often paid more attention to the piece selection policy rather than to the peer selection policy [11, 9, 12, 13]. Additionally, most of the research done so far has a limited point of view since it seldom takes into account the client's interactive behaviour nor differentiate whether the system being accessed is mostly *academic* or *non-academic*. Academic systems are those whose stored objects mostly refer to classes, lectures and tutorials, while non-academic systems are those whose most of the objects are movies, clips, songs and radio podcasts [10, 14, 15, 8, 5].

Within this context, this article proposes a simple algorithm to be used as basis for implementing peer selection policies of BitTorrent-like protocols, considering interactive scenarios. To this end, we analyze the client's interactive behaviour when accessing real multimedia systems. This analysis consists of looking into workloads of real content providers and assessing three important metrics, namely *temporal dispersion*, *spatial dispersion* and *object position popularity*. These metrics are then used as the main guidelines for writing the algorithm itself. To the best of our knowledge, our approach is unique since it is the very first to specially consider the client's interactive behaviour to derive an algorithm for peer selection policies targeted at interactive scenarios. Finally, the conclusion of this article is drawn with key challenges and future work in this subject matter.

The remainder of this text is organized as follows. In Section 2, we explain the P2P swarming paradigm and briefly review the operation of the BitTorrent protocol. Besides, we have a general discussion of the reciprocity principle, which plays a very important role when adapting

BitTorrent's peer selection policy to the on-demand streaming scenario. Section 3 presents the state-of-art proposals of this research field. Analysis and results all lie in Section 4. Therein we include a thorough discussion of the metrics used in the analysis and present the algorithm itself. At last, conclusions and future work are included in Section 5.

## 2. BASIS

### 2.1. Swarming systems and the BitTorrent protocol

A swarm is a set of peers concurrently sharing content of common interest. Content might be part of an object, an object or even a bundle of objects that are distributed together. The content is divided into pieces that peers upload to and download from each other. By leveraging resources provided by peers (i.e. users, clients, etc.), such as bandwidth and disk space, P2P swarming reduces the load on and costs to content providers and is certainly a scalable, robust and efficient solution for content replication [11, 8, 4].

BitTorrent is one of the most popular P2P swarming protocol. Objects are split into pieces (typically 32 – 256 kB in size) and each piece is split into blocks (typically 16 kB in size). Breaking pieces into blocks allows the protocol to always keep several requests (typically five) pipelined at once. Every time a block arrives, a new request is sent. This helps to avoid a delay between pieces being sent (i.e. *request response latency*), which is disastrous for transfer rates. Blocks are therefore the transmission unit on the network, but the protocol only accounts for transferred pieces. In particular, partially received pieces cannot be served by a peer, only complete pieces can [16, 17].

Peers who are still downloading content may serve the pieces they already have to others. *Leechers* are peers that only have some or none of the data, while *seeds* are peers that have all the data but stay in the system to let other peers download from them. Thus, seeds only perform uploading while leechers download pieces that they do not have and upload pieces that they have [16, 17].

To participate in the system (i.e. swarm), a peer firstly has to download a static file with the extension *.torrent* from an ordinary web server. The .torrent file has metadata (name, length, hashing information, etc.) of the wanted object. The peer is then able to contact a centralized entity called *tracker*. The main responsibility of this entity is to keep track of the participants of the distributed system and to provide the peer with a random list of other peers (both seeds and leechers) that already belong to the swarm. The tracker receives updates from peers periodically (typically at every 30 minutes) and when peers join or leave the swarm as well.

Each peer then attempts to establish bidirectional persistent connections (TCP connections) with a random set of peers (typically between 40 and 80) from the list just mentioned above, called its neighbourhood, but uploads data to only a subset of this neighbourhood. If the number of neighbours of a peer ever dips below 20, the node contacts the tracker again to obtain a list of additional peers it could connect to. The sequence of events just described is succinctly depicted in Figure 1 [2, 9].

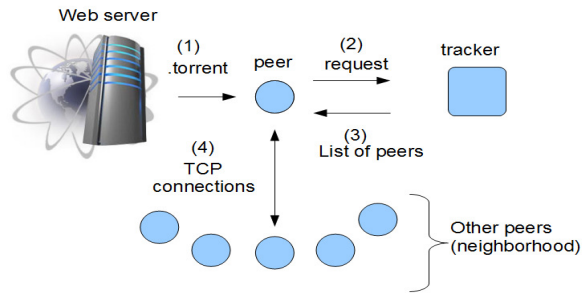

Figure 1. Sequence of events when a peer wants to join a swarm.

In a more detailed view, we have that each peer divides its upload capacity into a number of *uploads slots*. The decision to allocate an upload slot to a peer is termed an *unchoke* and there are two types of upload slots: *regular unchoke slots* and *optimistic unchoke slots*. Regular unchoke slots are assigned according to the *tit-for-tat policy*, i.e. peers prefer other peers that have recently provided data at the highest speeds. Still, each peer re-evaluates the allocation of its regular unchoke slots every *unchoke interval δ* (generally 10 seconds).

Different from the regular unchoke slots, the optimistic unchoke slots are assigned to randomly selected peers. Their allocation is re-evaluated every optimistic unchoke interval, which is generally set to 3δ (generally 30 seconds). Optimistic unchoke slots serve the purposes of (i) having peers discover other new, potentially faster, peers (nodes, users, clients, etc.) to unchoke so as to be reciprocated, and (ii) bootstrapping *newcomers* (i.e. peers with no pieces yet) in the system. Peers that are currently assigned an upload slot (whether regular or optimistic), from a node *p*, are said to be unchoked by node *p*; all the others are said to be choked by node *p* [2, 9]. In general, each peer limits the number of concurrent uploads (regular unchokes) to a small number, typically five and, even though seeds have nothing to download, they also limit the number of concurrent upload to typically five. As for the optimistic unchoke, there is usually a single current upload at a time [16, 17].

Lastly, each peer maintains its neighbourhood informed about the pieces it owns. The information received from its neighbourhood is used to request object pieces according to the local *rarest first* policy. This policy determines that each peer requests the pieces that are the rarest among its neighbours. The emergent effect of this policy is that less-replicated pieces get replicated fast among peers and possibly each peer obtains first the pieces that are most likely to interest its neighbours [17, 2, 9].

## 2.2. Reciprocity

### 2.2.1. Concepts

Reciprocity is one of the fundamental pillars supporting P2P systems. In essence, the principle of reciprocity states that participants must contribute to the system with resources, such as bandwidth or memory, in order to accomplish their tasks [18, 19].

Reciprocity mechanisms broadly classify in two types: *direct* and *indirect*. In the case of direct reciprocity, peers (users, clients, nodes, etc.) follow the principle of *I scratch your back and you*

International Journal of Computer Networks & Communications (IJCNC) Vol.5, No.5, September 2013

*scratch mine*. In the case of indirect reciprocity, the return may come from a peer other than the recipient, i.e. peer *A* provides content to *B* at a rate equal to that at which content is being provided to it by peer *C*, regardless of whether or not *B* and *C* are the same individual. In this case, peers follow the principle of *give and you shall be given*. Both direct and indirect reciprocity are used in P2P systems [18, 20]. This idea is depicted in Figure 2.

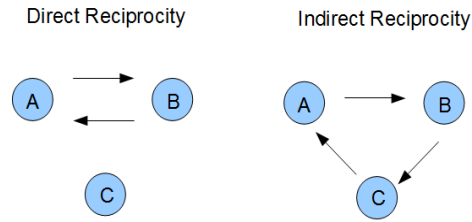

Figure 2. Direct and indirect reciprocities.

Thus, when a peer uses peer selection based on direct reciprocity, it will then upload to other peers that have recently uploaded to it at the highest rates. And when a peer uses peer selection based on indirect reciprocity, it will upload to other peers that have recently forwarded data to others at the highest rates [1].

**2.2.2. Reciprocity in BitTorrent**

As already mentioned, the BitTorrent' peer selection policy implements a direct reciprocity: tit-for-tat strategy. Under this strategy, peers prefer uploading to the peers that have recently contributed to them at the highest speeds. The emergent effect of this policy is that peers having higher uploads capacities also typically have higher download speeds [16], and hence more incentives for cooperation. Although these are desirable properties for the scenario of traditional file transfer, this policy has the two following drawbacks for on-demand streaming [2, 32, 33]:

i) direct reciprocity is less feasible in on-demand streaming, since it is difficult for younger peers (peers that arrived later) to reciprocate older peers (peers that arrived earlier). This is because peers that arrived later have made less progress on their downloads and have a playback position behind that of peers who have been in the system for longer; as a result, using tit-for-tat in on-demand streaming does not produce the same incentives as in traditional transfer;

ii) tit-for-tat was designed to maximize the download speed of peers. However, in on-demand streaming systems, peers gain little utility from having a download speed higher than the necessary to essentially preserve stream continuity. The use of tit-for-tat, in an on-demand streaming system with heterogeneous peers, may then lead to situations where, even if there is enough aggregate upload capacity to serve all system peers, not all peers experience a satisfactory QoS. This is because high-capacity peers may eventually receive download rates substantially higher than the object playback rate, while low-capacity peers may experience download rates that are too low to meet the on-demand streaming requirements.

From above, it becomes clear that the BitTorrent's peer selection policy result inadequate for on-demand streaming, even worse if we consider interactive scenarios [18, 1, 32, 33].

## 2.3. Aspects and Metrics

In this section, we succinctly debate on general aspects and metrics affecting the performance of P2P on-demand systems as well as its protocols. This surely helps to more adequately understand and clarify the workload analysis we carry out later in this text in order to better understand the client's behaviour when accessing real multimedia servers.

In general, we have the influence of the following six aspects affecting P2P VoD (video on-demand) systems [1, 6, 9]: (i) freeriding (i.e. the act of not sharing the upload capacity); (ii) malicious attacks; (iii) heterogeneous peer bandwidths (i.e. heterogeneity); (iv) presence of unconnectable peers (i.e. peers that do not possess a globally reachable address); (v) coexistence of peers doing VoD with peers doing traditional file transfer; and (vi) watching behaviour.

For the studies and analysis to happen herein, we conjecture that these same aspects may be taken into account when considering P2P on-demand systems. This is quite reasonable because a VoD system might be seen as a particular class of more general on-demand systems, in which the objects are specifically videos (i.e. a recording of moving pictures and sound).

With this in mind, the question that immediately follows is which of these aspects should be then evaluated when thinking of the design of peer selection policies for on-demand multimedia systems. The answer is that aspects (i), (iii) and (vi) are to be considered since they are the ones intrinsically related to the peer itself; the other aspects are mostly related to system and network infrastructure, security or service. This understanding is depicted in Figure 3.

As for performance metrics, there are plenty of them [1, 6, 9]. Just to name a few, we have: (i) continuity index (i.e. the ratio of pieces received before their deadline over the total number of pieces); (ii) startup delay (i.e. the time a client has to wait before playback starts); (iii) mean time to return (after an interruption); (iv) average number of interruptions (during the object playout); (v) total download time (of the object); (vi) bootstrap time (i.e. the time a client has to wait to obtain its first piece); (vii) link utilization; (viii) and fairness.

Similarly, the question that follows is which of the above metrics should be considered to design peer selection policies for P2P protocols. The answer is simple: all of the metrics listed above may be considered. This understanding is depicted in Figure 3. Nevertheless, we certainly recognize the high complexity involved with this thought, since the resulting values mostly come from simultaneous influences of the two policies that constitute the basis for P2P protocols: peer selection and piece selection. Moreover, it is intricate or maybe impossible to isolate each of these influences.

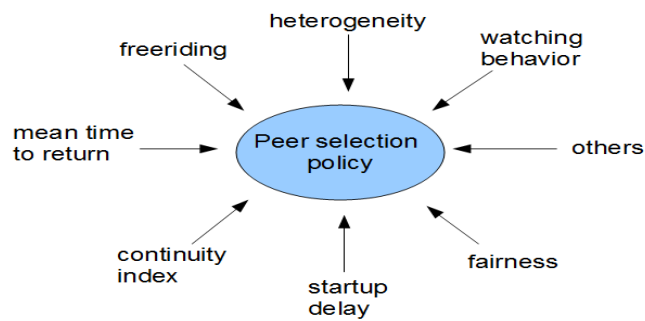

Figure 3. Aspects and metrics that may be considered for the design of peer selection policies.

## 3. STATE-OF-ART PEER SELECTION POLICIES

We recall that the peer selection policy determines how a peer (node, client, user) selects another peer to upload data to. In BitTorrent-like systems, this policy implicitly has the role of incentivizing peer cooperation. Hence, it is usually designed to favour good uploaders.

In what follows, we briefly discuss several important proposals of peer selection policies used by BitTorrent-like protocols targeted at VoD systems. Because of being recent and have been proven to be efficient [21, 22, 23, 2], these proposals certainly represent the state-of-art schemes of the literature. We mainly focus on how different the utilization of the regular and optimistic unchoke slots may be. Knowing these proposals certainly helps to better understand the workload analysis and the algorithm that appear in Section 4.

Shah et al. [21] proposes to use the optimistic unchoke slots more frequently. More precisely, every time a piece is played, peers perform a new optimistic unchoke. The idea is therefore to behave more altruistically. Problem is that frequent unchoke rounds may weaken incentives for cooperation; besides, a peer may compromise its own QoS by being too altruistic.

Mol et al. [22] dictates that, for the regular unchoke slots, each peer simply selects the nodes that have forwarded pieces received from it to other peers at the highest speed. Herein, we thus have the principle of indirect reciprocity clearly being applied. The optimistic unchoke slots are assigned in a similar way as in BitTorrent. However, as the reward given to peers is still similar to that of the tit-for-tat policy, it follows that heterogeneity can still lead to poor QoS for some peers, even in the presence of enough capacity to serve all.

Yang et al. [23] study six schemes to answer the question referred as the *peer request problem*: to which peer should a node send a request for a data piece, among all neighbours which have that piece? For instance, simply picking such a neighbour at random has the disadvantage that older peers receive more requests from many younger peers. Solving the peer request problem is therefore important to balance the load among the peers themselves and increase the likelihood of receiving the needed pieces before their deadlines. Both regular and optimistic unchoke slots are assigned in function of these schemes.

The first scheme of Yang et al. [23] is *Least Loaded Peer* (LLP): each node sends a request to the neighbour with the shortest queue size, among all those that have the needed data piece. The second scheme is *Least Requested Peer* (LRP): for each neighbour, each node counts how many requests it has sent to that peer and picks the one with the smallest count. The third scheme is *Tracker Assistant* (Tracker): the tracker sorts peers according to their arrival times. Whenever a node requests the list of available nodes from the tracker (e.g., upon arrival or when lacking peers due to peer departures), it will receive a list of nodes which have the closest arrival times to its own.

The fourth scheme of Yang et al. [23] is *Youngest-N Peers* (YNP): each node sorts its neighbours according to their age, where a peer's age can be determined from its join time (available at the tracker). YNP then randomly picks a peer among the $N > 1$ youngest peers which have the piece of interest and requests that piece from that neighbour. This approach tries to send requests to younger peers as they are less likely to be overloaded. The fifth scheme is *Closest-N Peers* (CNP): this is similar to YNP but instead sorts the neighbours based on how close they are to a node's own age, and then randomly picks from the $N$ closest age peers that have the needed piece. D'Acunto et al. [2] propose three schemes. They all make use of the indirect reciprocity principle: for the regular unchoke slots, each peer simply selects the nodes that have forwarded pieces received from it to other peers at the highest speed. In addition to that, these three schemes are also based on the idea of adjusting the number of upload slots a peer opens.

More precisely, the first scheme provides mathematical formulas to calculate the number of upload slots, as well as the number of optimistic unchoke slots. These formulas consider, for example, the peer's upload capacity as well as the video playback rate. The core of the second scheme of D'Acunto et al. [2] is to have peers dynamically adjusting the number of their optimistic slots to their current QoS. To measure the QoS, it is used the *sequential progress*, defined as the speed at which the index of the first missing piece grows. Finally, for the third scheme, they propose a mechanism where nodes give priority to newcomers when performing optimistic unchoke. In this way, newcomers will not need to wait too long before being unchoked for the first time.

From above, it seems that the recent proposals for peer selection policies are progressively evolving towards the explicit use of indirect reciprocity as well as the idea of continuously monitoring the QoS that high-capacity peers individually receive. In case the QoS is higher than necessary to preserve stream continuity, a number of altruistic decisions may be taken to favour low-capacity peers and thus provoke a more global satisfactory system QoS.

## 4. ANALYSIS

### 4.1. Workload characterization

In a sequential media workload, objects are completely and sequentially retrieved without any interruptions. A user (client) session refers to the period the client remains active in the system, i.e. making requests and receiving data. Note that, for this type of workload, the user session has just a single request: to start the content delivery. Also, if all peers (nodes) participating in the P2P system buffer all of the content received, then a newcomer may potentially select as neighbour any peer that has already received some content. This is because a peer that has already received some content is certain to receive the entire object. This scenario significantly facilitates the neighbour selection procedure and, consequently, the overall content sharing process.

On the other hand, in an interactive media workload, a user session consists of a number of independent requests to object segments. The object may be neither sequentially nor completely retrieved. Also, the time between consecutive requests may vary greatly. A number of characteristics of these workloads (see Table 1) may therefore have direct impact on content sharing among peers. For example, as a peer may retrieve only part of the object, there is less content to share with other peers. Besides, since a peer may request any object segment at any time, it is hard to predict if and when this peer will have the wanted content available to share with other peers. Peer selection strategies thus become very relevant since a poor selection may lead to a very poor content sharing.

Table 1. Content-sharing impacting characteristics.

| Characteristic | Definition |
|---|---|
| Request rate ($N$) | Rate at which requests from all users arrive during a period of time equal to the object duration. It is measured in requests per object. In other words it is the request rate normalized by the object duration. |
| Request Number ($R$) | The number of interactive requests executed by a user in a session. |
| Start position ($P_S$) | Position in object where the request starts, measured in time units. |
| End position ($P_E$) | Position in object where the request ends, measured in time units. |
| Request duration ($D_R$) | Amount of content transferred in a request, measured in time units. |
| Interaction type ($I_T$) | The type a request may be: pause, jump forwards, jump backwards, etc. |
| Duration of client inactivity ($1/R$) | Period in a session between the client's requests, measured in time units. |
| Jump distance ($J_D$) | Amount of content skipped in a jump, either forwards or backwards, measured in time units. |
| Session duration ($D_S$) | Period between the arrival of the first client's request and the end of the last client's request in a session, measured in time units. |

Our analysis considers interactive workloads from three different application domains. The domains are non-academic audio, non-academic video and academic video. For the academic domain, we look into workloads from eTeach [24], a video system from the University of Wisconsin-Madison, with 46,958 requests to announcement videos of up to five minutes as well as educational videos of 50-60 minutes. Additionally, we consider a three-year log from Manic [25], an educational video system from the University of Massachusetts with 25.833 requests.

For the non-academic domain, we analyze audio and video, typically under 10 minutes, from two major Latin America content providers. The first is UOL [26], with 5,385,822 requests to online radio objects and 1,453,117 requests to video objects. The second, for confidentiality denoted as ISP (Internet Service Provider), has 4,160,889 requests to online radio objects. All workloads just mentioned were also considered in the works of Costa et al. [27] and Rocha et al. [28], but with focus on interactive user access to exclusively client-server architectures.

We now use the characteristics mentioned in Table 1 to group our workloads in three interactivity profiles: High interactivity (HI): short request duration (under 20% of the object length), at least three requests in a session (i.e. $R > 2$), typical for academic videos; Low interactivity (LI): longer requests (at least 20% of multimedia object length), less than two requests in a session (i.e. $R < 2$), typical for non-academic audio and very short videos; Medium interactivity (MI): shorter request duration (under 20% of multimedia object length), less than three requests in a session (i.e. $R < 3$), typical for non-academic videos.

Figure 4 depicts the interactivity profile for three distinct real workloads. The graphs show the start position and the end position of each request. Requests are ordered by its start position. More precisely, Figure 4(a) shows the interactivity profile for a typical audio workload from UOL. This is a low interactivity profile, where most of the clients (users) retrieve all of the object. Figure 4(b) shows the interactivity profile for a typical video from Manic. This is a medium interactivity profile. Many clients also retrieve the entire object as there are many requests for object segments. Finally, Figure 4(b) shows a workload of a typical long video from eTeach. In such a high interactivity profile, the start positions and corresponding durations vary significantly.

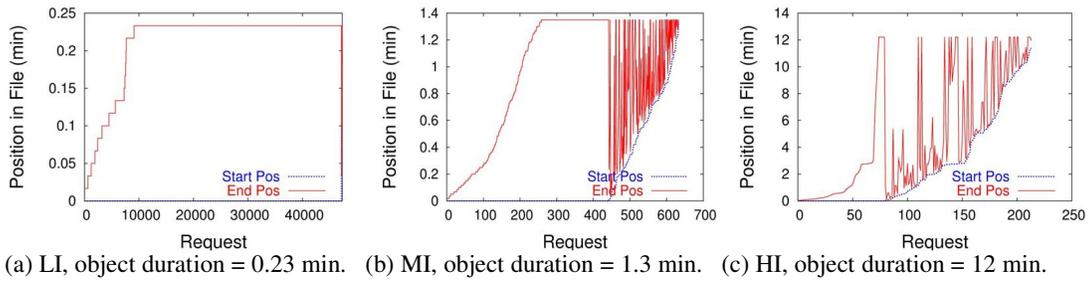

(a) LI, object duration = 0.23 min.  (b) MI, object duration = 1.3 min.  (c) HI, object duration = 12 min.

Figure 4. Interactivity profiles.

In total, we have 40 distinct workloads. Considering the interactivity profile, they are distributed as follows: 19 HI workloads; 8 MI workloads; and 13 LI workloads. On the other hand, considering the application domain, they are distributed as follows: 8 commercial audio workloads, 10 commercial video workloads; and 22 educational video workloads.

## 4.2. Dispersion and popularity

Section 2.3 shows potential aspects and metrics to be taken into account for the evaluation of peer selection strategies. Additionally, Table 1 brings several workload characteristics that may also have direct impact on content sharing between peers. Considering all of these evaluation parameters separately in our analysis is certainly prohibitive since it would become very intricate. Hence, to simplify we proceed as it follows.

First, we exclusively focus on the workload generated by a set of interactive clients (users) that are interested in sharing contents. Second, we conjecture that we may efficiently decide on the neighbour selection if we may evaluate the following values: (1) how much the requests diverge from each other considering the arrival times; (2) how much of the retrieved object segments may be effectively shared; (3) how often each object position is requested. So, to evaluate these three values, we define three distinct metrics: *temporal dispersion, spatial dispersion* and *object position popularity*.

Temporal dispersion is just another way to denote the request rate ($N$) and may be used to calculate the value (1). Low temporal dispersion implies high request rates and, consequently, more content available for sharing. On the other hand, high temporal dispersion implies low request rates and, consequently, less content available for sharing. By its turn, spatial dispersion refers to the amount of content two consecutive distinct users' requests have in common. It is used to calculate the value (2). As long as the requests have content in common, this content can be potentially shared. If this overlapping of consecutive requests increases, then the spatial dispersion decreases. On the other hand, if this overlapping of consecutive requests decreases, then spatial dispersion increases.

Figure 5 depicts the concept of both temporal and spatial dispersions, respectively. It shows four workloads especially devised for this example, each with ten requests for the same 25-minute object during the same time interval. More precisely, in Figure 5(a), we have a sequential workload. Each request retrieves the entire object. There is no spatial dispersion since all of the requests overlap. There are contents available for sharing. In Figure 5(b), spatial dispersion increases as the overlap between successive requests is reduced. As the overlap reduces and spatial dispersion increases, contents available for sharing are reduced. In Figure 5(c), spatial dispersion reduces as the overlap increases. Thus, there are contents available for sharing. Finally, in Figure 5(d), the interactive request rate $N$ increases and so decreases the temporal dispersion. There are contents available for sharing as well.

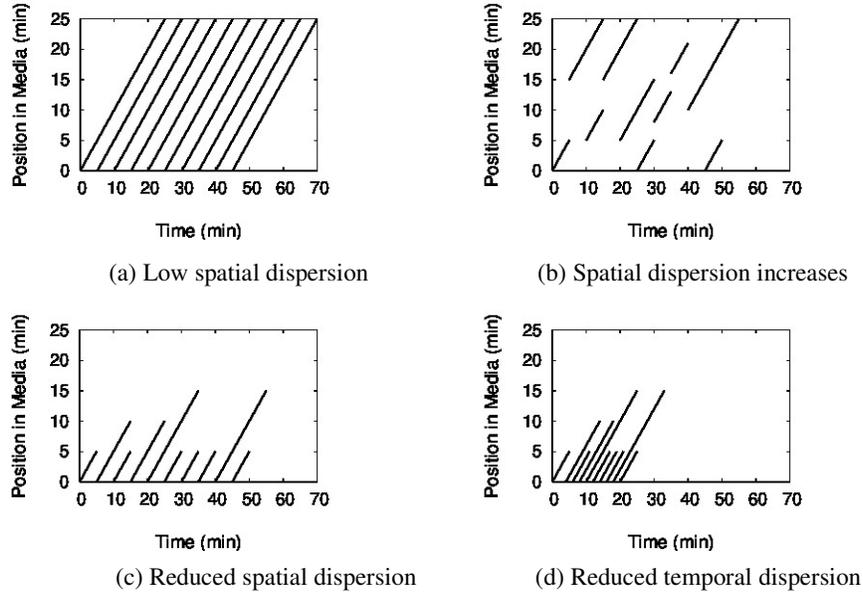

(a) Low spatial dispersion  
(b) Spatial dispersion increases  
(c) Reduced spatial dispersion  
(d) Reduced temporal dispersion  

Figure 5. Dispersion in workloads.

Measuring temporal dispersion is straightforward, as it is inversely proportional to the request rate $N$. To measure spatial dispersion we proceed as follows.

Let $Q_p$ to be the number of times an object position $p$ is requested by the system clients during the whole time of observation. Hence, $Q_p$ is the popularity of the object position $p$ (see Section 4.3.3). Any object position $p$, received by a peer, is potentially available to be shared with any other peers. It follows that the potential for content sharing $P$ is thus defined as it follows.

$$P = \sum_{p=0}^{T-1}(Q_p - 1) \qquad (1)$$

Now, to measure the spatial dispersion $D$, we compare the amount of content that is retrieved $M$ with the potential for content sharing $P$ as it follows.

$$D = 1 - \left(\frac{P}{M}\right) \qquad (2)$$

Note that $P < M$ and $D$ lies in the interval [0, 1). For example, if we have an object with a single position $p$ being requested 100 times, it follows that $Q_p = 100$ and $P = 99$. Since $M = 100$, it follows that $D = (1-(99/100)) = 0.01$. Still, note that the variable $Q_p$ is used to quantify the value (3) and is simple to be calculated. As already said, all we have to do is to count the number of times the object position is requested.

Additionally, we outline that, for ease of discussion, we consider the following intervals for a numerical dispersion categorization: below 0.1 (low spatial dispersion); (0.1, 0.5] (intermediate spatial dispersion); above 0.5 (high spatial dispersion).
To conclude this section, we outline that spatial dispersion, temporal dispersion and object position popularity are notably able to capture the influences coming from the aspects, metrics

International Journal of Computer Networks & Communications (IJCNC) Vol.5, No.5, September 2013

and workload characteristics presented in Figure 3 and Table 1, what makes the analysis we want much simpler and thus definitely viable. With this in mind, we have Figure 6. The aspects, metrics and characteristics, shown in Figure 3 and Table 1, have their corresponding influences grouped under the concept of spatial dispersion, temporal dispersion, popularity and/or system (i.e. related to system itself and network infrastructure, security or service). Note that a given aspect, metric or workload characteristic may simultaneously belong to more than one of those concepts.

### 4.3. Results

#### 4.3.1. Temporal dispersion

Most of our media workloads present high temporal dispersion. This is explained by the fact that in real multimedia servers the clients do not come in crowds since they individually have different needs and time availabilities for accessing the available contents.

For example, in academic systems, the contents being accessed mainly refer to video classes which may be seen at various distinct times during the day and night, since students are not confined to specific time schedules at all. Moreover, considering P2P swarming systems, the large measurement reported in [32] indicates that, for the vast majority of the datasets studied therein, around 40% to 70% of the swarms have only three or less peers and more than 70% of the swarms are of size smaller than ten [33]. These numbers reinforce the general belief that most of scenarios present workloads with high temporal dispersion.

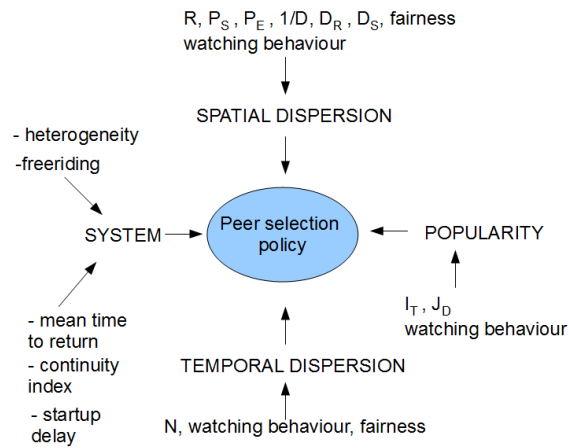

Figure 6. Aspects, workload characteristics and metrics affecting peer selection policies.

Nonetheless, a peer using the BitTorrent protocol stores all of the retrieved content (media) in local buffers. This buffering makes the time difference between clients' requests, i.e. the temporal dispersion, have little impact on content sharing. That is, it does not matter when the content is retrieved but that it is available for sharing when it is needed. This condition notably minimizes the impact of high temporal dispersion on content sharing.

On the other hand, if we have two workloads, with different request rates, soliciting data over the same amount of time, the one with higher request rate, i.e. with lower temporal dispersion, may potentially request more data and hence incur in lower spatial dispersion. That is, temporal dispersion may have some indirect impact on the spatial dispersion $D$ of a workload.

International Journal of Computer Networks & Communications (IJCNC) Vol.5, No.5, September 2013Figure 7 provides examples of the above situation. It refers to three of our workloads (for a 25-minute object). More precisely, in Figure 7(a), we have 10 requests at a rate of 4.55 and a spatial dispersion of 0.87. In Figure 7(b), we have 10 requests at a rate of 6.77 and a spatial dispersion of 0.87. Note that the overlapping between consecutive requests does not change as temporal dispersion varies. Finally, in Figure 7(c), we have a workload with 20 requests and a request rate of 9.09, twice the one observed in Figure 7(a). In this example, we have more requests and more content retrieved. The spatial dispersion reduces to 0.77. Note that the reduction is not due to temporal dispersion but to the increased number of requests instead.

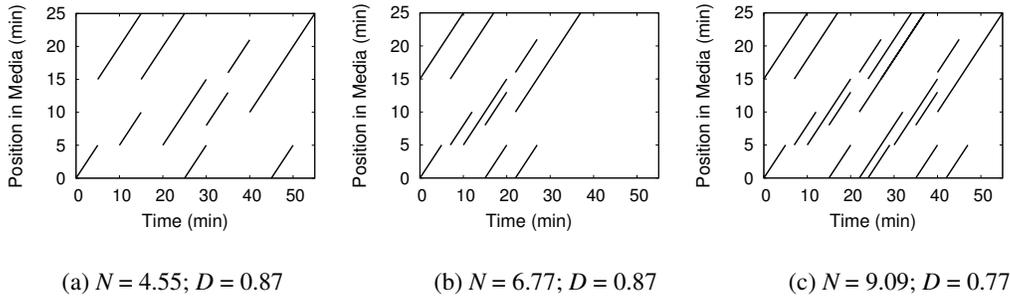

(a) $N = 4.55; D = 0.87$     (b) $N = 6.77; D = 0.87$     (c) $N = 9.09; D = 0.77$

Figure 7. Impact of temporal dispersion.

Thus, instead of having a quantitative assessment of the metric temporal dispersion, we simply consider the following general thought: *as low temporal dispersion may to some extent favour content sharing, due to its eventual impact on spatial dispersion, one should select as neighbours those peers with the lowest possible temporal dispersion.*

### 4.3.2. Spatial Dispersion

Figure 8(a) groups the workloads by their interactivity profile and the goal is to see how spatial dispersion varies for each of these profiles. The *y*-axis has the dispersion of each workload. The *x*-axis shows an arbitrary index of the workload within its interactivity profile. Note that LI profiles provide very low values of dispersion, MI profiles present intermediate values of dispersion, and HI profiles have high values of dispersion.

Figure 8(b) has the workloads grouped by the application domain and the goal is to see how the spatial dispersion varies within each of these domains. The *y*-axis brings the dispersion of each workload. The *x*-axis shows an arbitrary index of the workload within its domain of application: academic video, commercial video or commercial audio. Even though the dispersion may vary within a same domain, we may notice a tendency that allows us to conjecture that the dispersion is high for educational videos, medium for commercial videos, and low for commercial audio.

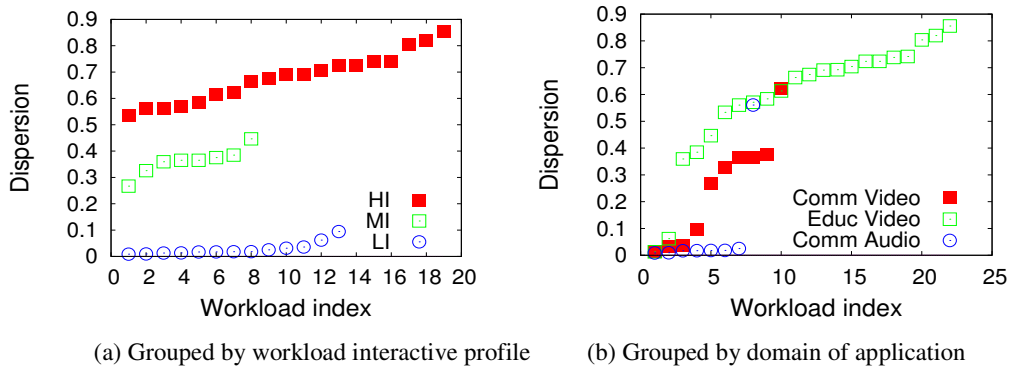

(a) Grouped by workload interactive profile     (b) Grouped by domain of application

Figure 8. Spatial dispersion in workloads.

International Journal of Computer Networks & Communications (IJCNC) Vol.5, No.5, September 2013

Figure 9 also groups the workloads by the application domain as done in Figure 8(b), but this time the goal is to see how the spatial dispersion varies within a same domain and considering distinct interactivity profiles. The *y*-axis has the dispersion of each workload and the *x*-axis an arbitrary index of the workload. More precisely, Figure 9(a) shows that most of the commercial audio workloads are of low-interactivity profile and have low dispersion. In Figure 9(b), we have that most of the workloads are of high-interactivity profile and have medium-to-high dispersion. Finally, from Figure 9(c), we see that the workloads are mainly from the low interactivity profile and medium interactivity profile. Note that, for the low interactivity profile, we have low dispersion and, for the medium interactivity profile, we have medium dispersion.

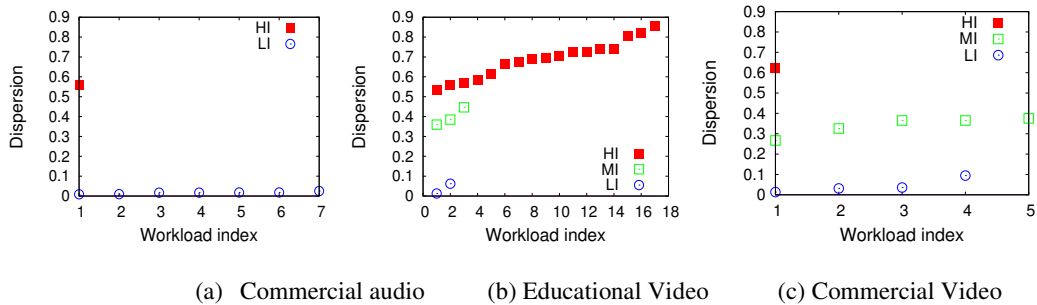

(a) Commercial audio    (b) Educational Video    (c) Commercial Video

Figure 9. Spatial dispersion in workloads of an application domain.

Figure 10 illustrates the spatial dispersion for the workloads grouped by the object length. The goal now is to see how this type of dispersion varies in function of this size within the application domains. The *y*-axis shows the dispersion of each workload and the *x*-axis has an arbitrary index of the workload. More precisely, Figure 10(a) clearly shows that most of the content (media) of less than five minutes has low dispersion. In other words, no matter the domain of application is, clients tend to retrieve shorter objects entirely. Figure 10(b) shows dispersion for 5-to-20 minute objects. Most of the objects falling in this category are from educational video domain and show a high dispersion. Finally, Figure 10(c) presents dispersion for objects of length longer than 20 minutes. In this case, we have that dispersion varies from medium to high

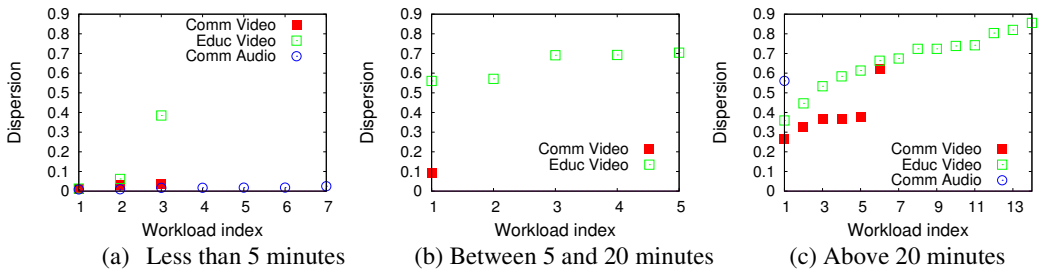

(a) Less than 5 minutes    (b) Between 5 and 20 minutes    (c) Above 20 minutes

Figure 10. Spatial dispersion in workloads considering objects of different lengths.

From the results just observed, we may conclude that: (i) the greater the interactivity is, the greater the corresponding dispersion is; (ii) the interactivity profile has more influence on the dispersion value than the application domain; (iii) the longer the object is, the greater the dispersion value is.

154

Thus, considering the spatial dispersion metric, we have the following general thought: *as low spatial dispersion makes more likely to have content sharing, one should always select as neighbours those peers with the lowest possible spatial dispersion.*

### 4.3.3. Object Position Popularity

The popularity of an object position $p$, denoted by $Q_p$, as previously defined in Section 4.2, is the number of times a position $p$ is requested by the clients, considering the whole time of observation. Figure 11 then shows the popularity of object positions for three workloads, one of each interactivity profile. The goal is to evaluate how the interactivity profile influences the object position popularity. The *x*-axis has each object (media) position in seconds. The *y*-axis brings the number of times each object position is solicited.

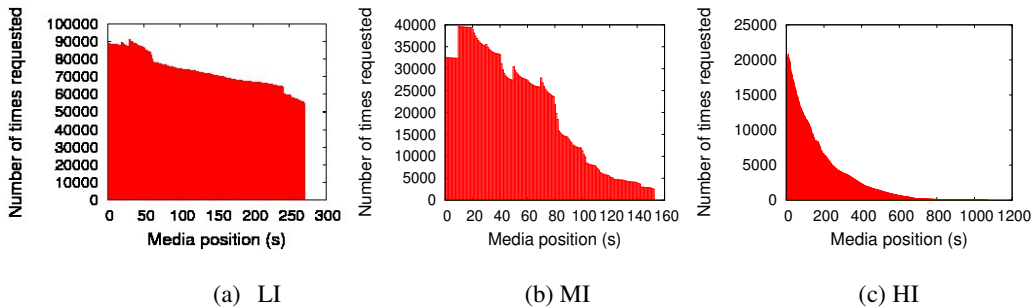

(a) LI    (b) MI    (c) HI

Figure 11. Popularity of object (media) positions.

From the results, for all interactivity profiles, the most popular object positions are often located in the beginning of the object. Also, the more interactive the workload is, the less uniform the popularity distribution is.

Thus, considering the object position popularity metric, we have the following general thought: *as the first positions of the object are more often requested, one should select as neighbours those peers that have already retrieved most of the first object positions, since this will result in more opportunities for content sharing.*

Lastly, it is important to clarify that the popularity definition is not related to the rarest-first definition used for piece selection in the traditional BitTorrent protocol. While the rarest-first definition refers to the number of times an object data unit is replicated among neighbours, position popularity specifically refers to the number of times an object position has been requested.

### 4.4. Algorithm

From the results and conclusions of last section, the design of efficient peer selection policies must dictate that the nodes to be selected as neighbours should help to reduce dispersion (temporal and spatial) as well take into account object position popularity, thus providing more opportunities for content sharing and, hence, optimizing the P2P system itself. To this end, we propose to observe the points that follow:

1) As in the traditional BitTorrent, a newcomer receives a list of possible neighbours from the tracker;
2) All retrieved contents are stored in local buffers. So there is no need for a peer to request the same contents twice;

3) Each node keeps a record of the number of times each object position is requested, i.e. the object position popularity. This record of popularity can be used as a basis to evaluate the potential for content sharing *P* in Equation 1 and, hence, to evaluate dispersion in Equation 2. Besides this record, each node also keeps a record of the buffer contents, as in the traditional BitTorrent protocol;
4) After receiving the list of possible neighbours, the peer exchanges its record of popularity with these possible neighbours. The record of buffer contents is also exchanged;
5) The peer can then assess the influence of each possible neighbour with respect to dispersion. This way, a peer can select its neighbours in such a way to have the neighbour set with minimum dispersion considering these steps:

Step 1: Let *C* be the set of peers received from the tracker and *S* the set of selected neighbours. Note that *S* is initially set empty;
Step 2: For each node *c* of *C*, evaluate the workload dispersion considering the peer set which has *c* and all of the nodes already in *S*, as follows:

Step 2.1: The node *c* that results in the lowest workload dispersion is selected as a new neighbour;
Step 2.2: This recently selected node is then removed from *C* and included in *S*;

Step 2.3: Stop when either: *C* is exhausted or *S* has the number of neighbours prescribed by the original BitTorrent protocol.

Finally, as complementary guidelines for the algorithm above, we still consider the points that follow:

a) For LI workloads, where most clients retrieve the object completely, it is preferable to select as neighbours those peers which have already started receiving data since they will more likely have the wanted contents available;
b) It is preferable to choose as neighbours those peers with highest request rates *N*, i.e. with the lowest temporal dispersion. Peers with higher request rates are more likely to fill their buffers fast and, thus, are more likely to have contents to share;
c) After forming the neighbour set as described above, the overall system upload and download capacities should be assessed. In case these capacities are satisfactory, nothing else needs to be done. In case these values are not satisfactory, the neighbourhood set needs to be re-evaluated: besides considering the dispersion (temporal and spatial) and position popularity metrics, the concept of indirect reciprocity should also be deployed when selecting the new neighbour in Step 2.1 above.

To conclude this section, we outline that the proposed algorithm is complete and rather robust since it is able to consider the three metrics herein defined in addition to the concept of indirect reciprocity. That is, the proposed algorithm indirectly covers the aspects, metrics and characteristics shown in Figure 3 and Table 1, besides the reciprocity concept, what certainly makes it applicable to the design of peer selection policies targeted at any kind of interactive scenario.

## 5. CONCLUSION AND FUTURE WORK

This article introduced an algorithm to be used as basis for implementing peer selection policies of BitTorrent-like protocols in interactive scenarios. To this end, we analyzed the client's interactive behaviour when accessing real multimedia systems. Our analysis considered workloads of real content providers and assessed three metrics herein defined: temporal dispersion, spatial dispersion and object position popularity. These metrics were then used the main guidelines for writing the proposed algorithm.

Among the most important results herein obtained, we may outline the following ones: (i) most of the media workloads analyzed presented high temporal dispersion. This is because the clients do not come in crowds since they individually have different needs and time availabilities for accessing the available contents. Nonetheless, since a peer using the BitTorrent protocol stores all of the retrieved content in local buffers, high temporal dispersion has little impact on content sharing; (ii) the spatial dispersion is a direct function of the client's interactivity level as well as of the object length; besides, it suffers more influences from the interactivity profile than the application domain; (iii) the most popular object positions are often located in the beginning of the object, and the more interactive the workload is, the less uniform the popularity distribution is; (iv) the proposed algorithm is rather complete in the sense that it is able to consider the three metrics herein defined besides the concept of indirect reciprocity.

Lastly, we believe future work in this field of study may specially include: (i) to quantify the optimization provided by the algorithm design insights revealed in this work. To this end, for example, it would be necessary to consider what has been revealed in this work along with what has already been proven efficient for piece selection policies [13, 8], resulting in novel proposals for BitTorrent-like protocols. These proposals should then guide the elaboration of models for extensive trace-driven simulations so that numerical observations could therefore be obtained; (ii) to study live streaming on the Internet to evaluate the eventual changes or adaptations that should occur in BitTorrent's peer selection and piece selection policies, respectively. This would certainly help the design of novel BitTorrent-like protocols targeted at the requirements of live streaming services [1, 29]; lastly, to analyse the influences of the personalization and adaptation to the client´s capability profile as well as of the Network Address Translation (NAT) service on the design of P2P streaming protocols [34, 35].

## REFERENCES

[1] D'Acunto, L., Chiluka, N., Vinkó, T. & Sips, H. (2013) "BitTorrent-like P2P approaches for VoD: A comparative study", *Computer Networks*, Vol. 57, No. 5, pp 1253 – 1276.
[2] D'Acunto, L., Andrade, J. & Sips, H. (2010) "Peer selection strategies for improved QoS in heterogeneous BitTorrent-like VoD systems", *IEEE International Symposium on Multimedia*, Taichung, Taiwan.
[3] Ramzan, N., Park, H. & Izquierdo, E. (2012) "Video streaming over P2P networks: Challenges and opportunities", *Signal processing: Image Communication*, Vol. 27, pp 401 – 411.
[4] Hoßfeld, T., Lehrieder, F., Hock, D., Oechsner, S., Despotovic, Z., Kellerer, W. & Michel, M. (2011) "Characterization of BitTorrent swarms and their distribution in the Internet", *Computer Networks*, Vol. 55, No. 5, pp 1197 – 1215.
[5] Hu, Chih-Lin, Chen, Da-You, Chang, & Chen, Yu-Wen. (2010) "Fair Peer Assignment Scheme for Peer-to-Peer File Sharing", *KSII Transactions on Internet and Information Systems*, Vol. 4, No. 5, pp 709 – 736.
[6] Bharambe, A., Herley, C. & Padmanabhan, V. (2006) "Analyzing and improving a BitTorrent network's performance mechanisms", *25th IEEE International Conference on Computer Communications*, Barcelona, Catalunya, Spain.
[7] Fun, X., Li, M., Ma, J., Ren, Y., Zhao, H. & Su, Z. (2012) "Behavior-based reputation management in P2P file-sharing networks", *Journal of Computer and System Sciences*, Vol. 78, No. 6, pp 1737 – 1750.
[8] Menasché, D., Rocha, A., Souza e Silva, E., Towsley, D. & Leão, R. (2011) "Implications of peer selection strategies by publishers on the performance of P2P swarming systems", *ACM SIGMETRICS Performance Evaluation Review*, Vol. 39, No. 3, pp 55 – 57.

[9]     Hoffmann, L. J., Rodrigues, C.K.S. & Leão, R. M. M. (2011) "BitTorrent-like protocols for interactive access to VoD systems", *European Journal of Scientific Research*, Vol. 58, No. 4, pp 550-569.
[10]    Varvello, M., Steiner, M. & Laevens, K. (2012) "Understanding BitTorrent: a reality check from the ISP's perspective", *Computer Networks*, Vol.56, No. 40, pp 1054 – 1065.
[11]    Menasché, D., Rocha, A., Souza e Silva, E., Leão, R., Towsley, D. & Venkataramani, A. (2010) "Estimating self-sustainability in peer-to-peer swarming systems", Performance Evaluation, Vol. 67, pp 1243 – 1258.
[12]    Chuntao, L., Huyin, Z. & Lijun, S. (2009) "Research and Design on Peer Selection Strategy of P2P Streaming", *5th International Conference on Wireless communications, networking and mobile computing*, Beijing, China.
[13]    Romero, P., Aomoza, F. & Rodrígues-Bocca, P. (2013) "Optimum piece selection strategies for a peer-to-peer video streaming platform", *Computers & Operations Research*, v. 40: 1289 – 1299.
[14]    Ye, L., Zhang, H., Li, F. & Su, M. (2010) "A measurement study on BitTorrent system", *International Journal of Communications, Network and System Sciences*, Vol. 3, pp 916 – 924.
[15]    Wu, T., Li, M. & Qi, M. (2010) "Optimizing peer selection in BitTorrent networks with genetic algorithms", *Future Generation Computer Systems*, Vol. 26, No. 8, pp 1151 – 1156.
[16]    Cohen, B. (2003) "Incentives build robustness in BitTorrent", *First Workshop on Economics of Peer-to-Peer Systems*, Berkeley, EUA.
[17]    Legout, A., Urvoy-Keller, G. & Michiardi, P. (2006) "Rarest first and choke algorithms are enough", *6th ACM SIGCOM Conference on Internet Measurement*, Rio de Janeiro, Brazil.
[18]    Menasché, D., Massoulié, L. & Towsley, D.  (2010) "Reciprocity and barter in peer-to-peer systems", *29th conference on Information communications*, San Diego, CA, USA.
[19]    Nowak, M. (2006) "Five rules for the evolution of cooperation", *Science*, Vol. 314, pp 1560 – 1563.
[20]    Nowak, M & Sigmund, K.  (2005) "Evolution of indirect reciprocity", *Nature*, Vol. 437, pp 1291–1298.
[21]    Shah, P. & Pâris, J.-F. (2007) "Peer-to-Peer multimedia streaming using BitTorrent", *IEEE International Performance, Computing, and Communications Conference – IPCCC*, New Orleans, EUA.
[22]    Mol, J., Pouwelse, J., Meulpolder, M., Epema, D. & Sips, H. (2008) "Give-to-Get: Free-riding-resilient video-on-demand in P2P systems", *SPIE MMCN*, San Jose, California, USA.
[23]    Yang, Y., Chow, A., Golubchik, L. & Bragg, D. (2010) "Improving QoS in Bittorrent-like VoD systems", *Proceedings of the IEEE INFOCOM*, San Diego, CA, USA.
[24]    http://eteach.cs.wisc.edu/index.html
[25]    RIPPLES/MANIC. http://manic.cs.umass.edu.
[26]    Universo Online. http://www.uol.com.br.
[27]    Costa, C., Cunha, I., Borges, A., Ramos, C., Rocha, M., Almeida, J., & Ribeiro-Neto, B. (2004) "Analyzing Client Interactive Behavior on Streaming Media Servers", *13th WWW Conf.*, New York, USA.
[28]    Rocha, M., Maia, M., Cunha, I., Almeida, J. & Campos, S. (2005) "Scalable Media Streaming to Interactive Users", *ACM MULTIMEDIA*, Singapore, Singapore.
[29]    Mol, J., Bakker, A., Pouwelse, J., Epema, D. & Sips, H. (2009) "The Design and Deployment of a BitTorrent Live Video Streaming Solution", *11th IEEE International Symposium on Multimedia*, San Diego, CA, USA.
[30]    Rocha, M. (2007) "Impactos da interatividade na escalabilidade de protocolos para mídia contínua", doctoral dissertation, *Universidade Federal de Minas Gerais*, Dept. Ciências da Computação, Belo Horizonte, MG, Brazil.
[31]    Liu, Y., Guo, Y. & Liang, C. (2008) "A survey on peer-to-peer video streaming systems". In Journal of Peer-to-peer Networking and Applications. Springer.
[32]    Hossfeld, T., Lehrieder, F., Hock, D., Oechsner, S., Despotovic, Z., Kellerer & Michel, M. (2011) "Characterization of BitTorrent swarms and their distribution in the Internet", *Computer Networks*, Vol.55, No. 5, pp 1197 – 1215.
[33]    de Souza e Silva, E., Leão, R., Menasché, D. & Rocha, A. (2013) "On the interplay between content popularity and performance in P2P systems", *10th International Conference, QEST 2013*, Buenos Aires, Argentina.

[34]  Zerkouk, M., Mhamed, A. & Messabih, B. (2013) "A user profile based access control model and architecture", *International Journal of Computer Networks & Communications (IJCNC)*, Vol. 5, No. 1, pp 171 – 181.

[35]  Masoud, M. Z. M (2013) "Analytical modelling of localized P2P streaming systems under NAT consideration", *International Journal of Computer Networks & Communications (IJCNC)*, Vol. 5, No. 3, pp 73 – 89.

**Authors**

Carlo Kleber da S. Rodrigues received the B.Sc. degree in Electrical Engineering from the Federal University of Paraiba in 1993, the M.Sc. degree in Systems and Computation from IME in 2000, the D.Sc. degree in System Engineering and Computation from the Federal University of Rio de Janeiro in 2006. Currently he is Military Assessor of the Brazilian Army in Ecuador, Professor at the Polytechnic School of Army in Ecuador, and Professor at the University Center UniCEUB in Brazil. His research interests include the areas of computer networks and multimedia systems.

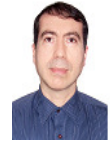

Marcus Vinícius de Melo Rocha received the B.Sc. degree in Computer Sciences from the Federal University of Minas Gerais in 1984, the M.Sc. degree in Sciences from the Federal University of Minas Gerais in 2002, and the D.Sc. degree in Sciences from the Federal University of Minas Gerais in 2007. Currently works with internet and streaming infrastructure in Minas Gerais State Assembly (MG/Brazil). His research interests include the areas of computer networks and streaming systems.

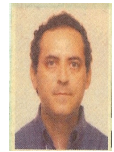